\documentclass[aps,prb,twocolumn,superscriptaddress,floatfix]{revtex4}
\usepackage{graphicx}
\usepackage{amssymb}

\begin{document}

\title{Phase Diagram of Ba$_{1-x}$K$_{x}$Fe$_2$As$_2$}

\author{S. Avci}
\affiliation{Materials Science Division, Argonne National Laboratory, Argonne, IL 60439-4845, USA}
\author{O. Chmaissem}
\affiliation{Materials Science Division, Argonne National Laboratory, Argonne, IL 60439-4845, USA}
\affiliation{Physics Department, Northern Illinois University, DeKalb, IL 60115, USA}
\author{D. Y. Chung}
\author{S. Rosenkranz}
\affiliation{Materials Science Division, Argonne National Laboratory, Argonne, IL 60439-4845, USA}
\author{E. A. Goremychkin}
\affiliation{Materials Science Division, Argonne National Laboratory, Argonne, IL 60439-4845, USA}
\affiliation{ISIS Neutron and Muon Source, Rutherford Appleton Laboratory, Didcot, OX11 0QX, United Kingdom}
\author{J.-P. Castellan}
\author{I. S. Todorov}
\author{J. A. Schlueter}
\author{H. Claus}
\affiliation{Materials Science Division, Argonne National Laboratory, Argonne, IL 60439-4845, USA}
\author{A. Daoud-Aladine}
\author{D. D. Khalyavin}
\affiliation{ISIS Neutron and Muon Source, Rutherford Appleton Laboratory, Didcot, OX11 0QX, United Kingdom}
\author{M. G. Kanatzidis}
\affiliation{Materials Science Division, Argonne National Laboratory, Argonne, IL 60439-4845, USA}
\affiliation{Department of Chemistry, Northwestern University, Evanston, IL 60208-3113, USA}
\author{R. Osborn}
\affiliation{Materials Science Division, Argonne National Laboratory, Argonne, IL 60439-4845, USA}
\email{SAvci@anl.gov}

\begin{abstract}
We report the results of a systematic investigation of the phase diagram of the iron-based superconductor, Ba$_{1-x}$K$_{x}$Fe$_2$As$_2$, from $x = 0$ to $x = 1.0$ using high resolution neutron and x-ray diffraction and magnetization measurements. The polycrystalline samples were prepared with an estimated compositional variation of $\Delta x \lesssim 0.01$, allowing a more precise estimate of the phase boundaries than reported so far. At room temperature, Ba$_{1-x}$K$_{x}$Fe$_2$As$_2$ crystallizes in a tetragonal structure with the space group symmetry of $I4/mmm$, but at low doping, the samples undergo a coincident first-order structural and magnetic phase transition to an orthorhombic (O) structure with space group $Fmmm$ and a striped antiferromagnet (AF) with space group $F_{c}mm^{\prime}m^{\prime}$. The transition temperature falls from a maximum of 139\,K in the undoped compound to 0\,K at $x = 0.252$, with a critical exponent as a function of doping of 0.25(2) and 0.12(1) for the structural and magnetic order parameters, respectively. The onset of superconductivity occurs at a critical concentration of $x = 0.130(3)$ and the superconducting transition temperature grows linearly with $x$ until it crosses the AF/O phase boundary. Below this concentration, there is microscopic phase coexistence of the AF/O and superconducting order parameters, although a slight suppression of the AF/O order is evidence that the phases are competing. At higher doping, superconductivity has a maximum $T_c$ of 38\,K at $x = 0.4$ falling to 3\,K at $x = 1.0$. We discuss reasons for the suppression of the spin-density-wave order and the electron-hole asymmetry in the phase diagram.
\end{abstract}

\date{\today}

\maketitle

\section{Introduction}

There is now an extensive body of research into the origin of superconductivity in the iron-based superconductors demonstrating the importance of the subtle interplay of their electronic properties with crystalline structure [1,2]. These compounds all contain a common structural motif, namely a square planar net of iron atoms tetrahedrally coordinated with pnictogens or chalcogens producing Fe$_2X_2$ layers, in which $X$ = As or Se/Te in the highest $T_c$ compounds. They are separated by buffer layers comprising, for example, rare earth oxides in the so-called 1111 systems, such as those based on LaFeAsO, or alkaline earths in the so-called 122 systems, such as those based on BaFe$_2$As$_2$. Like the cuprate superconductors, the buffer layers can act as charge reservoirs, controlling the carrier concentration and inducing superconductivity in the iron planes by the introduction of aliovalent dopants, \textit{e.g.}, LaFeAsO$_{1-x}$F$_x$ [3,4] and Ba$_{1-x}$K$_{x}$Fe$_2$As$_2$ [5,6], but it is also possible to dope the pnictogen or chalcogen sites, \textit{e.g.}, BaFe$_2$As$_{2-x}$P$_x$ [7,8], or the iron planes themselves by substituting other transition metal ions, \textit{e.g.}, BaFe$_{2-x}$Co$_x$As$_2$ and BaFe$_{2-x}$Ni$_x$As$_2$ [9,10]. 

In this article, we report on a systematic investigation of the phase diagram of Ba$_{1-x}$K$_{x}$Fe$_2$As$_2$ from $x = 0$ to $x = 1.0$ using high resolution neutron and x-ray diffraction combined with bulk characterization. In spite of the diversity of doping strategies employed to modify the superconducting properties of the iron-based superconductors, their phase diagrams show remarkable similarities. There is typically an undoped ``parent" compound that is antiferromagnetic rather than superconducting[1]. These are fully compensated metals, whose electronic properties are dominated by multiple iron-derived $d$-bands near the Fermi level with approximately equal concentrations of hole and electron carriers [11]. The Fermi surfaces consist of quasi-two-dimensional cylinders, with two or three hole pockets at the Brillouin zone centers and two electron pockets at the $M$-points on the zone boundaries, \textit{i.e.}, along the direction of the nearest-neighbor iron-iron bonds [12]. All the Fermi surface pockets have similar radii in the undoped compounds, making their electronic structure particularly susceptible to magnetic instabilities resulting from a nesting of the disconnected hole and electron Fermi surfaces [13]. Since the nesting wavevector corresponds to the antiferromagnetic wavevector observed by neutron diffraction [14-16], it is plausible that the magnetism can be explained by a purely itinerant model of spin density waves, and \textit{ab initio} density functional theory does indeed predict the correct magnetic structure [17], although there is an ongoing debate about the strength of electron correlations [18,19].

The magnetic structure breaks the tetragonal symmetry with an in-plane wavevector of either (0,$\pi$) or ($\pi$,0) in the unfolded Brillouin zone with one iron atom per unit cell. With a finite magnetoelastic coupling, this would induce an orthorhombic structural transition, which is usually observed to occur at the same temperature as magnetic order in the parent compounds [20,21]. However, the addition of both hole and electron charge carriers through chemical substitution suppresses both transitions. A variety of scenarios are possible in Ginzburg-Landau treatments of the magnetoelastic coupling [22,23]. The two phase transitions could be first- or second-order and occur simultaneously or separately. In most of the iron-based compounds, the two transition temperatures split with doping [24] and there is a report of a split transition in Ba$_{1-x}$K$_{x}$Fe$_2$As$_2$ as well [25]. However, this is inconsistent with our previously reported neutron diffraction data, which shows unambiguously that they are coincident and first-order for all $x$ before they are both suppressed at $x\lesssim0.3$ [6]. Unusually, the two order parameters, magnetic and structural, are proportional to each other, apparently indicating a biquadratic coupling that is usually only observed at a tetracritical point [26,27], not over an extended range of compositions. Possible explanations for this observation will be discussed in the conclusions.

Superconductivity emerges before the complete suppression of the antiferromagnetic/orthorhombic (AF/O) phase and coexists at low-doping levels. The nature of the competition between AF/O order and superconductivity is a central question in understanding iron-based superconductivity [28,29]. There were earlier reports based on local probes that, in Ba$_{1-x}$K$_{x}$Fe$_2$As$_2$, the coexistence region is characterized by a mesoscopic phase separation into AF/O and superconducting droplets [30,31], but our previously reported diffraction data are only consistent with a microscopic phase coexistence [6], a conclusion since supported by muon spin rotation ($\mu$SR) experiments [32], suggesting that the earlier reports may be due to compositional fluctuations within the samples.

\begin{figure}[!t]
\centering
\includegraphics[width=0.6\columnwidth]{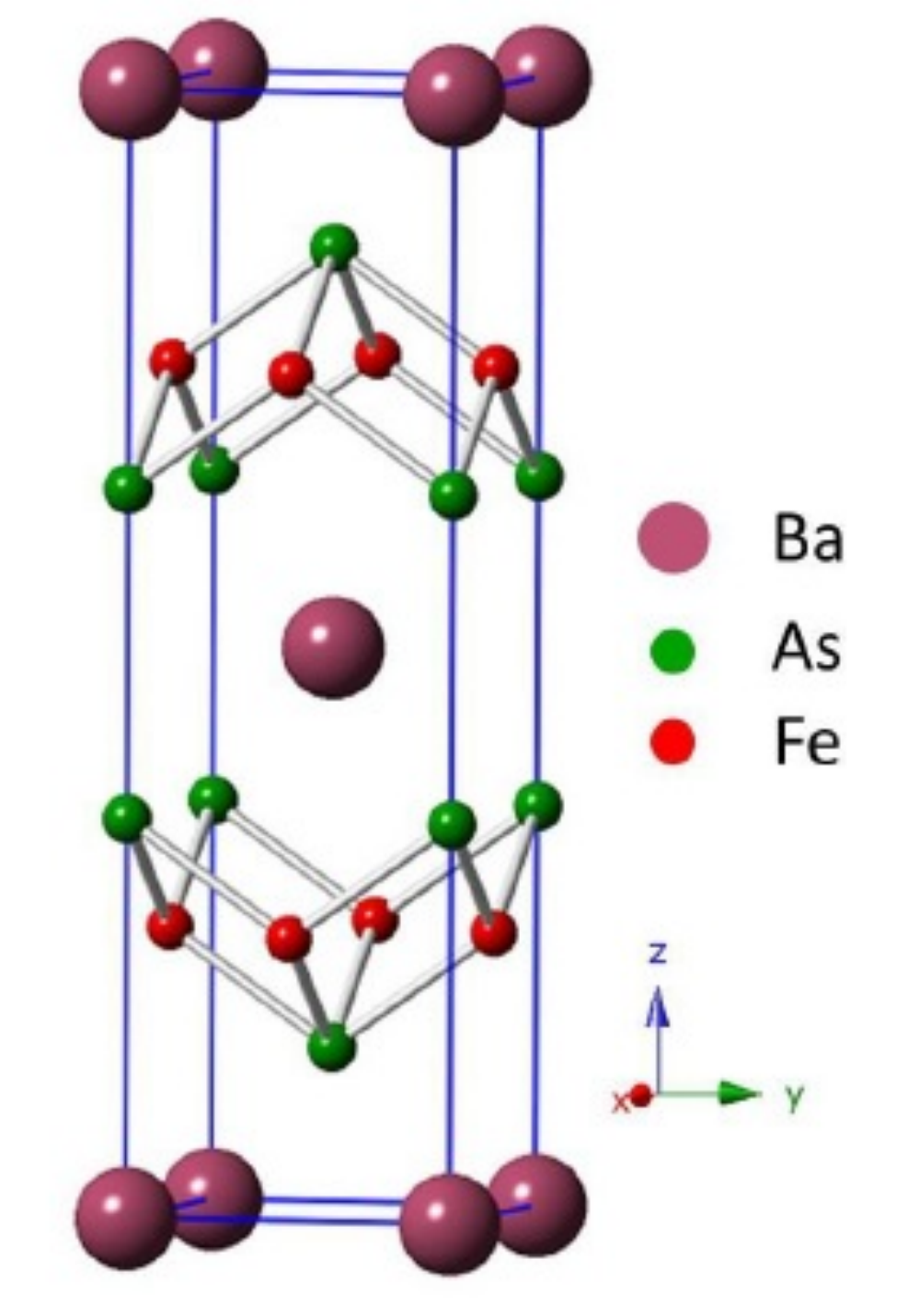}
\vspace{-0.15in}
\caption{(Color online) Structure of BaFe$_2$As$_2$, which crystallizes in a tetragonal ThCr$_2$Si$_2$-type structure with the space group symmetry of $I4/mmm$. Potassium substitutes onto the barium sites.
\label{Fig1} }
\vspace{-0.2in}
\end{figure}

One of the main reasons for studying Ba$_{1-x}$K$_{x}$Fe$_2$As$_2$ is that superconductivity extends up to much higher hole-doping levels, with 0.5 holes/Fe atom, than in the electron-doped superconductors produced by transition metal substitutions. In the case of BaFe$_{2-x}$Co$_x$As$_2$, superconductivity vanishes at only 0.12 electrons/Fe atom [10]. Furthermore, the maximum $T_c$ with hole-doping is 38\,K, significantly higher than the maximum $T_c$ of $\sim25$\,K obtained with electron doping. This electron-hole asymmetry in the phase diagram has been attributed to an enhanced Fermi surface nesting in the hole-doped compounds, consistent with angle-resolved photoemission spectroscopy (ARPES) data and band structure calculations [33]. This explanation is also supported by the evolution of resonant spin excitations, which become incommensurate due to the mismatch in hole and electron Fermi surface volumes when $T_c$ starts to fall [34]. On the other hand, there is also a strong correlation between $T_c$ and internal structural parameters such as the Fe-As-Fe bond angles [35,36]. These are known to have an influence on the band structure and the degree of moment localization, but their role in optimizing superconductivity and the implications for the gap symmetry is a matter of debate [37,38]. 

There have been two previous reports of the doping dependence of this series in addition to our own brief report, which are all in qualitative agreement [6,36,39]. At room temperature, all members of the Ba$_{1-x}$K$_{x}$Fe$_2$As$_2$ series crystallize in a tetragonal structure with the space group symmetry of $I4/mmm$ (Fig. 1), while low-doped samples also exhibit a low temperature phase transition to an orthorhombic structure with space group $Fmmm$ [5]. Superconducting samples at higher doping have a maximum $T_c$ of 38\,K and remain tetragonal at all measured temperatures down to 1.7\,K. However, there are significant discrepancies in the published reports concerning the critical dopant concentrations defining the onset of superconductivity and the suppression of the AF/O phase, with the latter varying from $x\sim0.3$ [36] to $\sim0.4$ [39]. As already mentioned, there have also been disagreements about the nature of the competition between the three ordered phases at low doping. We believe that these discrepancies are due to uncertainties in the actual composition of the synthesized samples, since it is well-known that potassium is particularly volatile. Controlling the inhomogeneity to within acceptable limits in order to improve the accuracy of the various phase boundaries has been a key goal of this work and we estimate that we have been able to make samples in which $\Delta x<0.01$. We have performed neutron and x-ray diffraction studies of the magnetic and structural order using high-resolution powder diffractometers so that the systematic variation of the lattice parameters and internal structural parameters can be used to estimate the degree of uncertainty in the average composition and its variation within the samples.

\begin{table*}[!t]
\begin{center}
\begin{tabular}{cccccccccc}
\hline
$x$ & $x$ & $x$ & $x$ & $T_c$ (K) & $T_N$ (K) & $T_N$ (K) & $T_s$ (K) & Magn. Moment ($\mu_B$) & $\delta \times10^3$ \\
& (fitted) & (nominal) & (ICP) & & (magnetization) & (neutron) &  &  &  \\
\hline
0 &  & 0 & 0 &  & 139(1) & 139.0(1) & 138.17(6) & 0.756(36) & 3.92(4) \\
0.1 & 0.097 & 0.1 & 0.094(2) &  & 136(1) & 136.5(3) & 136.02(8) & 0.741(21) & 3.68(3) \\
0.125 & 0.126 & 0.125 & 0.114(2) &  & 130(2) &  & 128.29(6) & 0.697(29) & 3.49(4) \\
0.15 & 0.15 & 0.15 & 0.139(1) & 4 & 122(2) & 122.1(3) & 122.09(7) & 0.702(21) & 3.35(4) \\
0.175 & 0.172 & 0.175 & 0.159(2) & 10 & 113(2) & 113.9(1) & 112.1(6) & 0.683(24) & 3.14(5) \\
0.2 & 0.202 & 0.2 & 0.184(2) & 17 & 100(2) & 102.0(1) & 102.00(2) & 0.652(46) & 2.76(7) \\
0.21 & 0.209 & 0.24 &  & 18 &  & 96.0(1) & 96.0(3) & 0.610(32) & 2.59(3) \\
0.22 & 0.225 & 0.22 &  & 23.5 &  & 93.97(1) & 93.93(1) & 0.550(22) & 2.20(9) \\
0.24 & 0.237 & 0.26 &  & 26 &  & 79.9(1) & 80.0(2) & 0.572(29) & 2.00(3) \\
0.25 & 0.249 & 0.24 &  & 28.5 &  & 74.9(1) & 74.8(8) & 0.456(22) & 1.43(8) \\
0.28 &  & 0.28 &  & 34 &  &  &  &  & \\
0.3 &  & 0.3 & 0.312(4) & 36 &  &  &  &  & \\
0.4 &  & 0.4 &  & 38 &  &  &  &  & \\
0.5 &  & 0.5 & 0.476(8) & 34 &  &  &  &  & \\
0.7 &  & 0.7 & 0.675(3) & 20 &  &  &  &  & \\
0.9 &  & 0.9 & 0.892(1) & 7 &  &  &  &  & \\
1 &  & 1 & 1.00(2) & 3 &  &  &  &  & \\
\hline
\end{tabular}
\end{center}
\caption{Structural and magnetic phase diagram of Ba$_{1-x}$K$_{x}$Fe$_2$As$_2$. The nominal value of x represents the starting composition given by the Ba/Fe ratio. The fitted value is determined by smoothing the variation in the a-axis lattice parameter from $0.1\leq x\leq0.25$ using a power law function. The first column is the value of $x$ used in the text and figures. The superconducting transition temperature is determined from magnetization measurements. $T_N$ and $T_s$ are determined by magnetization and neutron diffraction data as described in the text. The magnetic moments and orthorhombic order parameters, $\delta=(a-b)/(a+b)$, are determined from the low-temperature Rietveld refinements. 
\label{Table1}}
\end{table*}

In this article, we present the results of Rietveld refinements for the entire series and use this analysis along with bulk measurements to produce a comprehensive magnetic and structural phase diagram that provides insight into the nature of the phase competition that underlies iron-based superconductivity. Our results show that there is a steeper decrease in $T_c$ and hence a narrower region of phase coexistence of the AF/O order with superconductivity than previous reports. After a description of our experimental results, we combine our findings with results reported in the literature on the electron-doped BaFe$_{2-x}$Co$_x$As$_2$ series in order to elucidate the origin of the electron-hole asymmetry in the phase diagram.

\section{Experimental Details}
The synthesis of homogeneous single phase Ba$_{1-x}$K$_{x}$Fe$_2$As$_2$ samples is known to be particularly delicate due to unfavorable kinetics, high vapor pressures, and a significant difference in the chemical reactivity of K and Ba with FeAs that may result in stabilizing other binary by-products. For this work, the synthesis and properties of our samples were optimized by the systematic examination of all reasonable combinations of reaction parameters, \textit{e.g.}, purity of the starting materials, reaction containers, temperature, and duration of heating, etc. Our final samples were produced according to the following procedure: Handling of all materials was performed in a nitrogen-filled glove box. Raw materials (BaAs/KAs/Fe$_2$As) were prepared by heating elemental mixtures at 400$^\circ$C, 600$^\circ$C, and 850$^\circ$C, respectively. The stoichiometric mixture of these starting precursors for a desired composition was thoroughly ground to ensure uniform and homogeneous and subsequently annealed at 1050$^\circ$C in a Nb tube sealed in a quartz tube. The closed metal tubes are needed to eliminate any chemical loss that may otherwise result from the evaporation of K and As. A large number of high quality samples were synthesized covering the full phase diagram $0\leq x\leq1$ with special emphasis given to the $0.1\leq x\leq0.25$ range in which the potassium content was incremented in very small amounts with $\Delta x=0.025$. The deliberate synthesis of samples with finely tuned K content was necessary in order to carefully investigate the rapid suppression of magnetism with increasing K and to elucidate the nature of phase coexistence with superconductivity within the same sample. Samples with coarse K increments would otherwise lead to inconclusive results.

Initial characterization of the samples was performed by x-ray diffraction, magnetization measurements, and Inductively Coupled Plasma (ICP) elemental analysis. For select samples, neutron powder diffraction experiments were performed on the HRPD ($x=0$, 0.1, 0.125, 0.15, 0.175, 0.2, 0.21, 0.24, 0.3, 0.5 and 1) and Wish diffractometers ($x=0.22$ and 0.25) at the ISIS Pulsed Neutron and Muon Source (Rutherford Appleton Laboratory) and on beamline 11-BM, ($x=0.28$) at the Advanced Photon Source (Argonne National Laboratory). The resolution, $\Delta d/d$, at 2\,\AA is 0.001 for HRPD and 0.002 for Wish. For diffraction experiments, the samples were sealed under vacuum for shipment and re-opened just before the measurements in a helium environment to prevent air exposure. The nuclear and magnetic structures together with interatomic bond-lengths and bond-angles were determined by the Rietveld refinement technique using the comprehensive General Structure Analysis System (GSAS) software suite [40] and the associated graphical user interface (EXPGUI) [41]. Minute traces of no more than 0.5-3\% by weight of FeAs and Fe$_2$As impurity phases were observed in some of the samples, too small to affect the analysis.

Our neutron diffraction results show that the synthesis methods has reduced the compositional uncertainty significantly, which we estimate to be $\Delta x\lesssim0.01$. In previously reported phase diagrams, it is not totally clear whether the potassium contents were nominal or actual measured values. Because of its volatility, it is always necessary to add excess potassium, so the eventual stoichiometry is largely governed by the Ba/Fe ratio. Thus, any perceived discrepancy with other work, such as Ref.  [39] is probably due to differences in the handling and control of the volatile potassium and arsenic constituents. In this article, we have used a number of methods to characterize the sample compositions, including direct measurements of the stoichiometry from ICP analysis. To produce the final compositions used in our phase diagrams, we started with the nominal $x$ values determined from the starting Ba/Fe ratio and then smoothed the variation in the a-axis lattice parameter from $0.1\leq x\leq0.25$ using a power law function (see Table 1). In most cases, the agreement with the nominal value was better than 0.005, with just two samples requiring a significant shift of 0.02 and 0.03, respectively. In all cases, the adjustments also improved the consistency of other measurements, such as the variations in transition temperatures and order parameters. The first column of Table 1 shows the value of $x$ that we have used in the text and figure labels, derived by rounding the fitted values to the nearest 0.005, although the fitted values were used in the plots and in numerical analyses of the doping dependence, \textit{e.g.}, the fit of $T_c$ \textit{vs} $x$. For $x\geq0.28$, where the precise composition is not so critical, we have used the nominal values.

\begin{figure}[!b]
\centering
\includegraphics[width=\columnwidth]{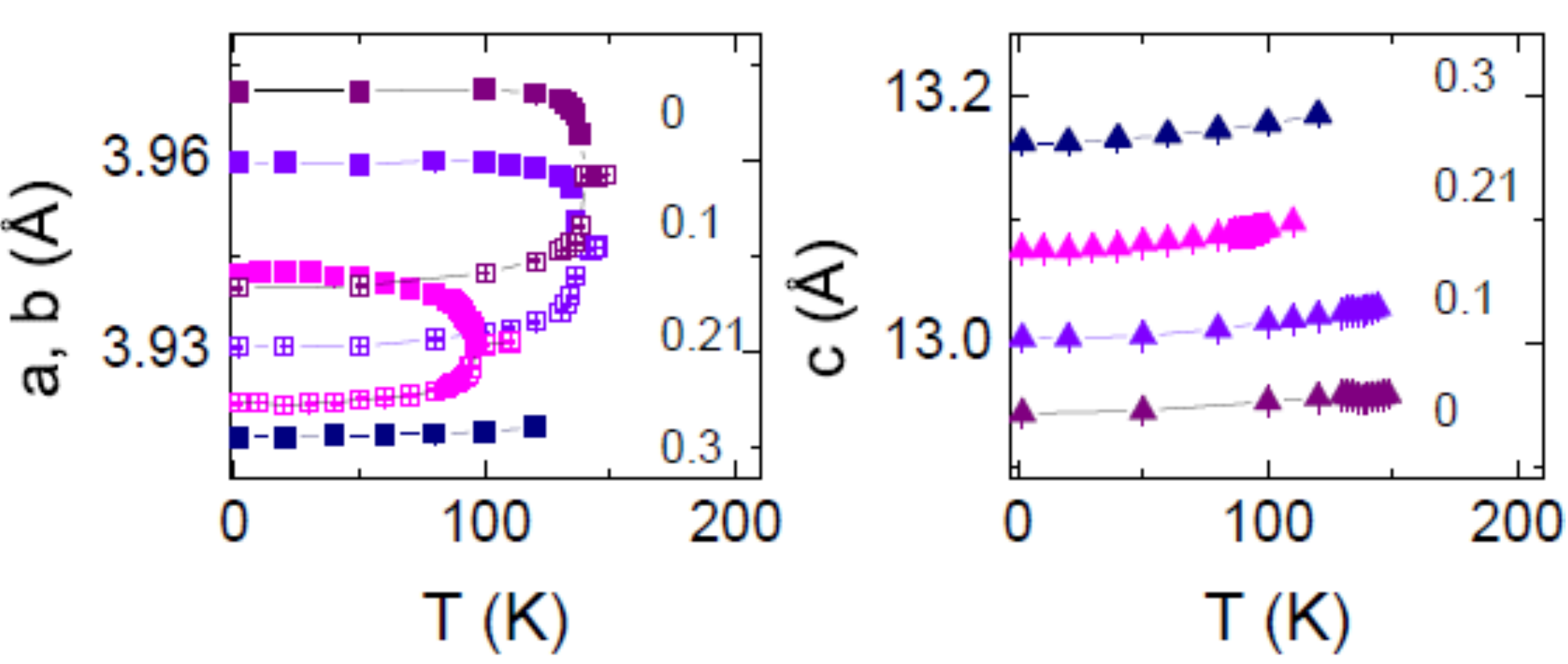}
\vspace{-0.15in}
\caption{(Color online) Variation of lattice constants, $a$, $b$ and $c$, with temperature for $x = 0$, 0.1, 0.21 and 0.3. The lattice constants, $a$ and $b$, in the orthorhombic phase are divided by $\sqrt{2}$.
\label{Fig2} }
\vspace{-0.2in}
\end{figure}

We are also able to monitor fluctuations in composition within a single sample because HRPD has sufficiently high resolution that it is sensitive to distributions of the lattice parameter and internal strains caused by compositional gradients and even particle size broadening [42]. The diffraction peaks from the (220) reflection shown in Fig. 1 of Ref.  [6] show that there is no change in the linewidths and lineshapes of the diffraction peaks from the undoped compound up to $x=0.24$, consistent with a high-degree of compositional homogeneity ($\Delta x\lesssim0.01$). The regular spacing between the peak positions is in agreement with the fixed steps in $x$.

\section{Results}
\subsection{Structural Phase Diagram}

\begin{figure}[!b]
\centering
\includegraphics[width=\columnwidth]{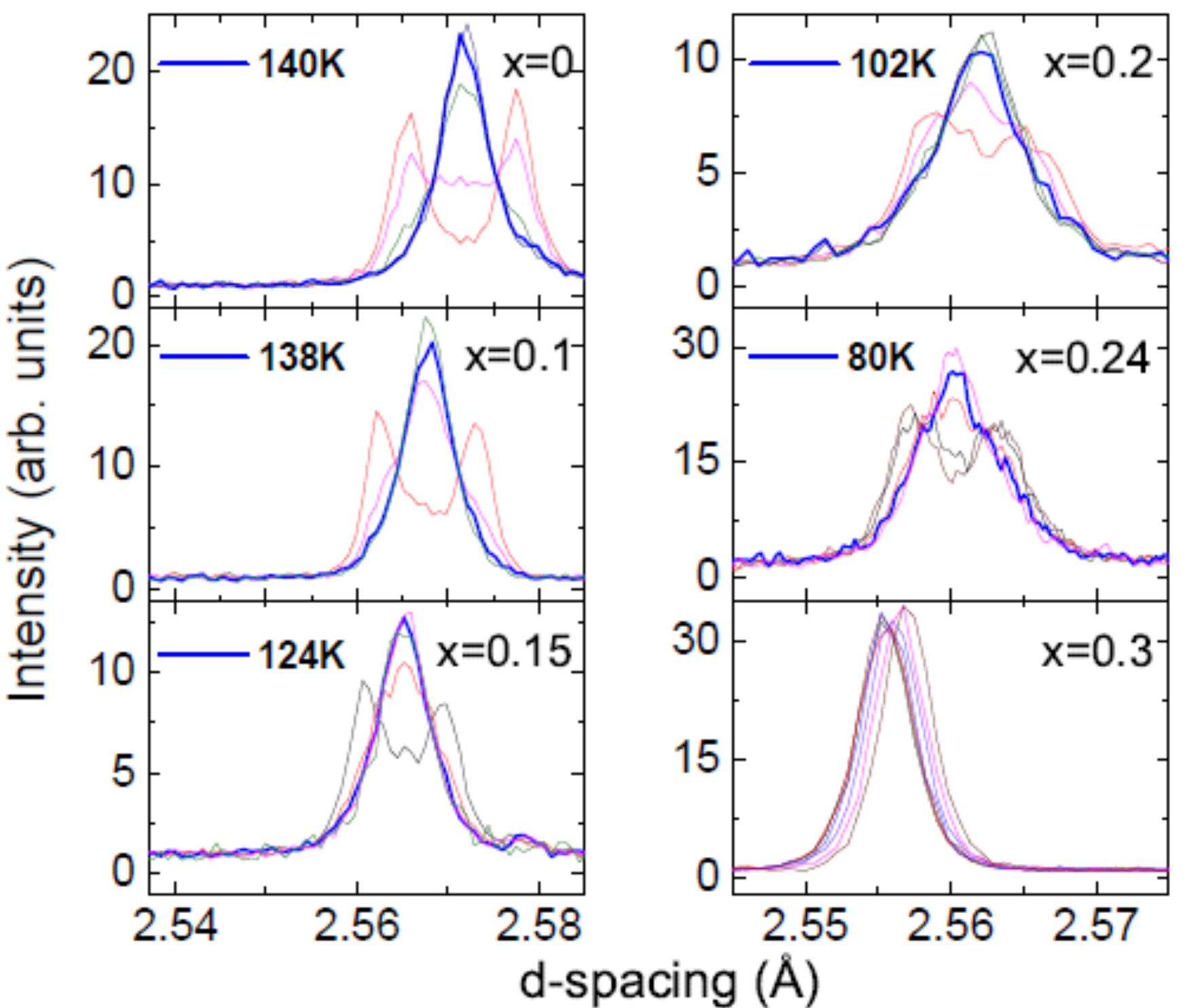}
\vspace{-0.15in}
\caption{(Color online) Temperature dependence of the (110) peak in the vicinity of structural transition temperature $T_s$. The bold curve shows the peak at the estimated $T_s$. For $0\leq x\leq0.24$, the other curves show the peak at intervals of 2\,K  around $T_s$. For $x=0.3$, the peak is shown in 20\,K intervals between 1.7\,K and 120\,K.
\label{Fig3} }
\vspace{-0.2in}
\end{figure}

In the undoped BaFe$_2$As$_2$ compound, there is a structural phase transition at 139\,K from the tetragonal ThCr$_2$Si$_2$-type structure with the space group symmetry $I4/mmm$ to an orthorhombic $\beta$-SrRh$_2$As$_2$-type structure with space group $Fmmm$ [5]. The structure of both the tetragonal and orthorhombic phases of Ba$_{1-x}$K$_{x}$Fe$_2$As$_2$ can be described as a stack of edge-sharing Fe$_2$As$_2$ layers separated by layers of (Ba,K) ions (Fig. 1). The (Ba,K) ions occupy crystallographic positions that are tetrahedrally coordinated with four arsenic anions.

With potassium doping, the tetragonal to orthorhombic structural transition temperature, $T_s$, decreases until it is fully suppressed for $x>0.25$. In Fig. 2, we show the lattice parameters as a function of temperature for $x=0$, 0.1, 0.21 and 0.3. Below $x=0.3$, a significant orthorhombic splitting of the basal plane $a$ and $b$ lattice parameters is observed. The evolution of the orthorhombic order parameter defined by $\delta=(a-b)/(a+b)$ is discussed later. Although the transitions appear to be continuous, we observed small but sharp volume anomalies at all the structural phase transitions (Fig. 3 in Ref. [6]) showing that they are weakly first-order in character over the entire phase diagram.

\begin{figure}[!t]
\centering
\includegraphics[width=0.8\columnwidth]{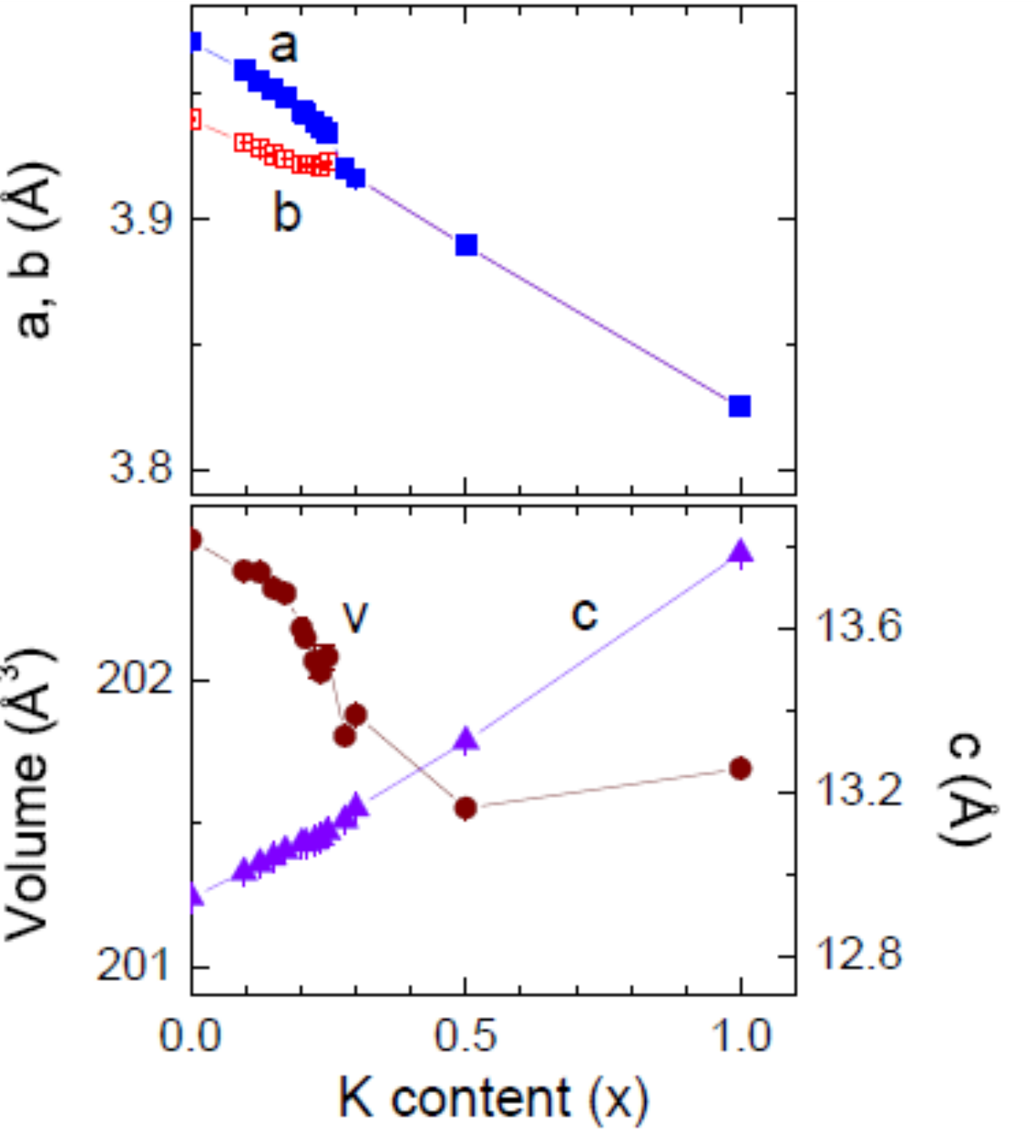}
\vspace{-0.15in}
\caption{(Color online) Variation of lattice constants and volume in Ba$_{1-x}$K$_x$Fe$_2$As$_2$ with $x$ at 1.7\,K. Solid lines are guide to the eye. The lattice constants ($a$ and $b$) and the volume in the orthorhombic phase are divided by $\sqrt{2}$ and 2, respectively, from those for the $Fmmm$ space group.
\label{Fig4} }
\vspace{-0.2in}
\end{figure}

The structural transition temperatures in Table 1 were determined by fitting a power law, $\delta\propto(T_s-T)^\beta/T_s$, close to the transition temperature, yielding exponents mostly in the range $\beta\sim0.13$ to 0.2. Apart from $x=0$, where $\beta=0.129(3)$, in reasonable agreement with Ref.  [27], the exponents will be modified by compositional fluctuations and so are not reliable estimates of the critical behavior. The exponents at $x=0.21$ and 0.24 were anomalously high ($\beta=0.25$ and 0.30, respectively), which could reflect a slightly greater degree of compositional variation within those samples.  As a check on these values of $T_s$, we used the peak profiles close to the transition temperature. Fig. 3 shows the temperature dependence of the tetragonal (110) Bragg peak for $x=0$, 0.1, 0.15, 0.2, 0.24 and 0.3 in the vicinity of the structural transition. This reflection splits into two (022) and (202) orthorhombic peaks below $T_s$. Close to the transition, the two peaks cannot be resolved but we can determine $T_s$ from the temperature dependence of the full width at half maximum (FWHM) as they merge. In the high-temperature phase, the FWHM of $\sim0.0037(3)$\AA is independent of the composition. These two methods of determining $T_s$ agreed within the errors.

\begin{figure}[!t]
\centering
\includegraphics[width=\columnwidth]{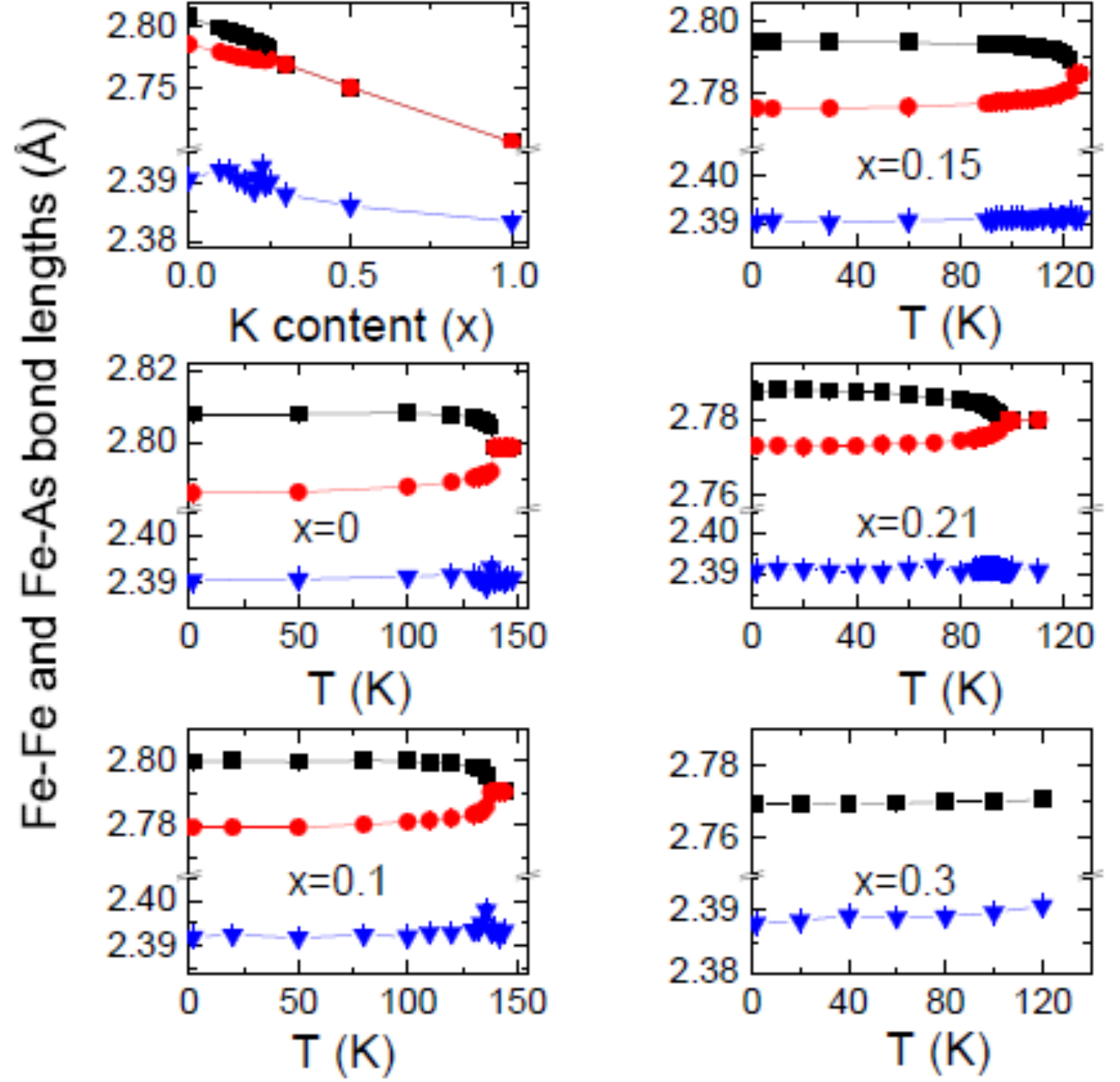}
\caption{(Color online) Variation of Fe-Fe and Fe-As bond lengths with $x$ at 1.7\, K and with temperature for different K substitutions. Blue triangles represent the Fe-As bonds. Black square and red circle symbols represent the Fe-Fe bond lengths merging at $T_s$. Solid lines are guides to the eye.
\label{Fig5} }
\vspace{-0.2in}
\end{figure}

\begin{table*}[!t]
\begin{center}
\footnotesize
\begin{tabular}{cccccccccccc}
\hline
$x$ & $a$(\AA) & $b$(\AA) & $c$(\AA) & Vol(\AA$^3$) & $z$As & Fe-As(\AA) & Fe-Fe(\AA) & Fe-Fe(\AA) & As-Fe-As($^\circ$) & As-Fe-As($^\circ$) & As-Fe-As($^\circ$) \\
\hline
0 & 5.6157(2) & 5.5718(2) & 12.9424(4) & 404.970(45) & 0.35375(3) & 2.3905(1) & 2.8078(1) & 2.7859(1) & 108.071(3) & 108.718(3) & 111.648(2) \\
0.1 & 5.5997(1) & 5.5587(1) & 13.0031(4) & 404.755(36) & 0.35405(3) & 2.3919(1) & 2.7998(1) & 2.7793(1) & 108.356(7) & 108.961(7) & 111.110(15) \\
0.125 & 5.5940(2) & 5.5551(2) & 13.0243(5) & 404.745(52) & 0.35405(3) & 2.3918(2) & 2.7970(1) & 2.7775(1) & 108.438(8) & 109.011(8) & 110.977(16) \\
0.15 & 5.5890(2) & 5.5517(2) & 13.0404(5) & 404.634(46) & 0.35399(3) & 2.3905(1) & 2.7945(1) & 2.7758(1) & 108.464(3) & 109.013(3) & 110.947(3) \\
0.175 & 5.5842(2) & 5.5492(2) & 13.0563(6) & 404.598(55) & 0.35386(3) & 2.3901(2) & 2.7921(1) & 2.7746(1) & 108.522(9) & 109.037(9) & 110.864(17) \\
0.2 & 5.5767(4) & 5.5460(4) & 13.0736(9) & 404.353(82) & 0.35372(5) & 2.3885(1) & 2.7883(2) & 2.7730(2) & 108.578(5) & 109.030(5) & 110.815(4) \\
0.21 & 5.5750(1) & 5.5462(2) & 13.0749(4) & 404.288(36) & 0.35404(2) & 2.39073(4) & 2.7875(1) & 2.7731(1) & 108.678(2) & 109.102(2) & 110.640(2) \\
0.22 & 5.5706(5) & 5.5461(4) & 13.0803(12) & 404.126(112) & 0.3543(7) & 2.3925(6) & 2.7855(3) & 2.7733(3) & 108.798(21) & 109.155(21) & 110.47(4) \\
0.24 & 5.5672(2) & 5.5449(2) & 13.0888(4) & 404.051(39) & 0.35394(3) & 2.3894(2) & 2.7836(1) & 2.7724(1) & 108.750(7) & 109.078(7) & 110.592(14) \\
0.25 & 5.5636(4) & 5.5476(4) & 13.1027(10) & 404.157(90) & 0.35398(6) & 2.3900(5) & 2.7810(2) & 2.7732(2) & 108.846(17) & 109.073(16) & 110.50(3) \\
0.3 & 3.9165(2) & 3.9165(2) & 13.1614(5) & 201.877(23) & 0.35383(4) & 2.3878(3) & 2.7694(2) & 2.7694(2) & 109.090(28) &  & 110.237(14) \\
0.5 & 3.8893(2) &  & 13.3242(6) & 201.554(20) & 0.35376(4) & 2.3859(3) & 2.7501(1) &  & 109.615(11) &  & 109.185(21) \\
1 & 3.8251(2) &  & 13.7846(5) & 201.691(40) & 0.35314(4) & 2.3833(4) & 2.7047(1) &  & 110.855(12) &  & 106.708(23) \\
\hline
\end{tabular}
\end{center}
\caption{Results of Rietveld refinements for Ba$_{1-x}$K$_x$Fe$_2$As$_2$ from neutron powder diffraction data collected on HRPD at 1.7\,K. For $x<0.3$, the space group is $Fmmm$, in which $a\neq b$, there are two inequivalent Fe-Fe bond distances and three inequivalent As-Fe-As bond angles. For $x\geq 0.3$, the space group is $I4/mmm$, in which $a=b$, there is one Fe-Fe bond distance and two inequivalent As-Fe-As bond angles
\label{Table2}}
\end{table*}%

Fig. 4 shows the evolution of the lattice parameters and unit cell volume at 1.7\,K as a function of $x$. Barium substitution by the smaller potassium cations reduces the in-plane $a$- and $b$-lattice parameters while significantly lengthening the out-of-plane $c$-axis. However, the c-axis enhancement is not large enough to fully compensate for the shrinking basal plane axes and the unit cell volume gradually decreases in magnitude upon increasing the K content until $x\sim0.5$. The reason for the non-monotonic behavior of the unit cell volume at high dopant levels is not understood and will require further investigation.

The behavior of the Fe-Fe and Fe-As interatomic distances (at 1.7\,K) are presented in Fig. 5. The Fe-Fe distances mimic the in-plane lattice parameters both as a function of K content and of temperature. The similarity in behavior is explained by the fact that Fe atoms occupy special rigid positions along the long edges of the lattice. Six As-Fe-As bond-angles can be identified in each FeAs$_4$ tetrahedron. In the tetragonal $I4/mmm$ structure, these angles can be grouped into two independent angles: two equivalent angles, $\alpha_1$, and four equivalent smaller ones, $\alpha_2$. In the orthorhombic $Fmmm$ structure below $T_N$, the angle $\alpha_1$ remains unaffected but the angle $\alpha_2$ splits into two pairs of equivalent angles $\alpha^{\prime}_2$ and $\alpha^{\prime\prime}_2$ with an angular separation of $\sim0.5-0.6^\circ$. A sketch showing the different angles is displayed in Fig. 6. As shown in the same panel, the angle $\alpha_1$ increases linearly and continuously throughout the whole phase diagram. However, starting from BaFe$_2$As$_2$, the angles $\alpha^{\prime}_2$ and $\alpha^{\prime\prime}_2$ increase with increasing K until reaching a critical composition below $x\sim0.3$ beyond which the structural transitions are suppressed and the two angles merge into $\alpha_2$, which continues to increase with higher K contents. The refined values of the lattice parameters, bond lengths and bond angles at 1.7\,K are shown in Table 2.

\begin{figure}[!b]
\centering
\includegraphics[width=\columnwidth]{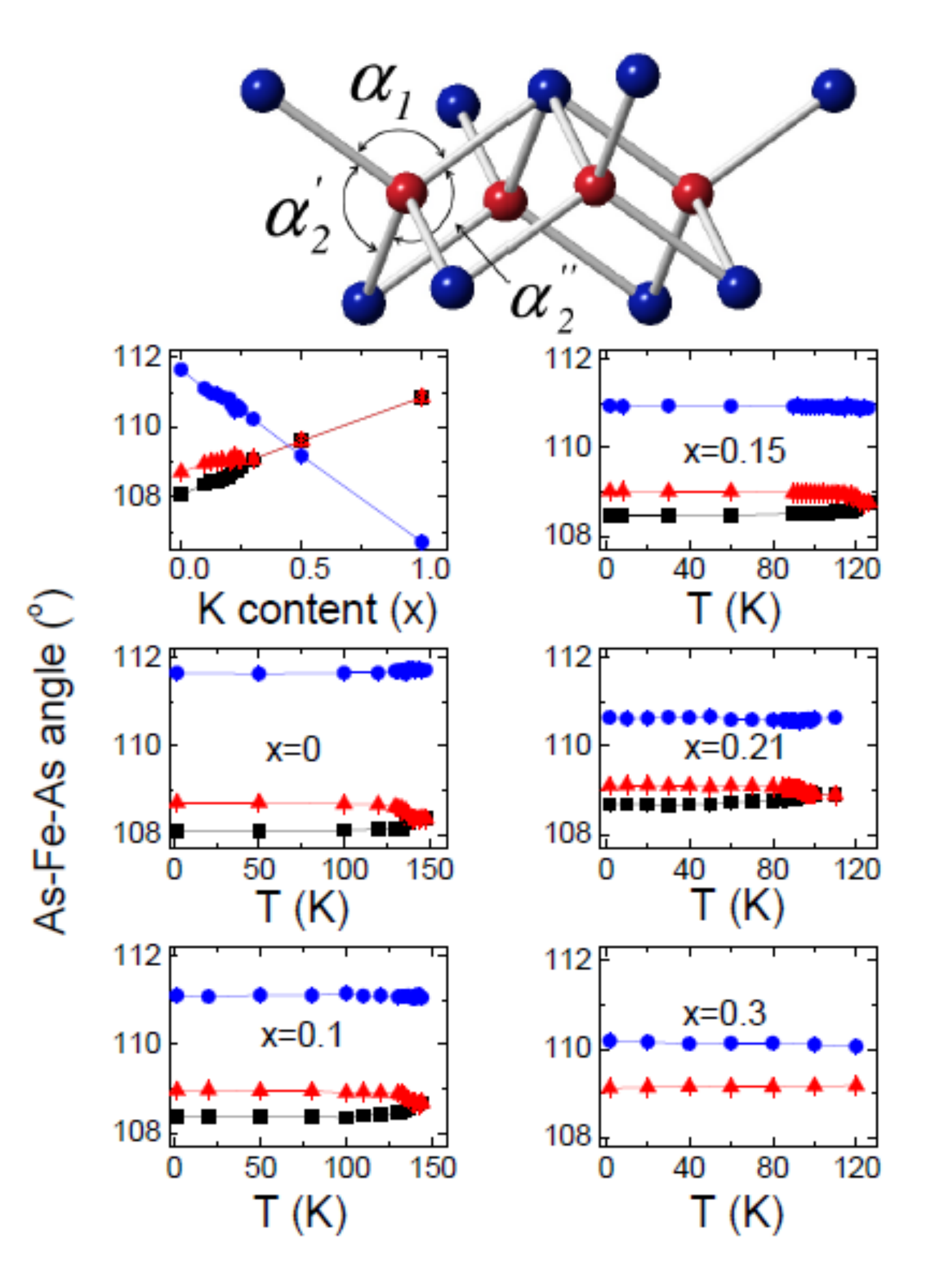}
\caption{(Color online) Variation of As-Fe-As bond angles with $x$ at 1.7\, K and with temperature for different K substitutions. The top panel shows $\alpha_1$, $\alpha^\prime_2$ and $\alpha^{\prime\prime}_2$ in orthorhombic setting. Blue circles represent $\alpha_1$, red triangles and black squares represent $\alpha^\prime_2$  and $\alpha^{\prime\prime}_2$ merging into one $\alpha_2$  at $T_s$. Solid lines are guides to the eye.
\label{Fig6} }
\vspace{-0.2in}
\end{figure}

\subsection{Magnetic Phase Diagram}

Neutron powder diffraction data reveal the presence of weak magnetic Bragg reflections that appear below the structural phase transition for all of the orthorhombic samples. The magnetic peaks shown in Fig. 7, located at 2.45\,\AA and 3.43\,\AA, were indexed as 121 and 103 in agreement with the widely reported antiferromagnetic spin density wave (SDW) ground state [15,16,39,43]. As with the structural transitions, the antiferromagnetic transition (N\'eel) temperatures as a function of doping were determined by power-law fits to temperature variation of the magnetic moment (see Table 1). These coincide with the orthorhombic transition for all values of $x$. It is important to realize that the structural and magnetic orders are identified in the same measurements, the first from the splitting of the nuclear Bragg peaks and the second by the intensity of the magnetic Bragg peaks. This means that the conclusion that the two transitions are coincident does not depend on the accuracy of the thermometry. As shown in Ref.  [6], the two order parameters determined in this way are directly proportional to each other at all temperatures over the entire phase diagram, an unusual result that we will discuss in more detail in the discussion.

\begin{figure}[!t]
\centering
\includegraphics[width=\columnwidth]{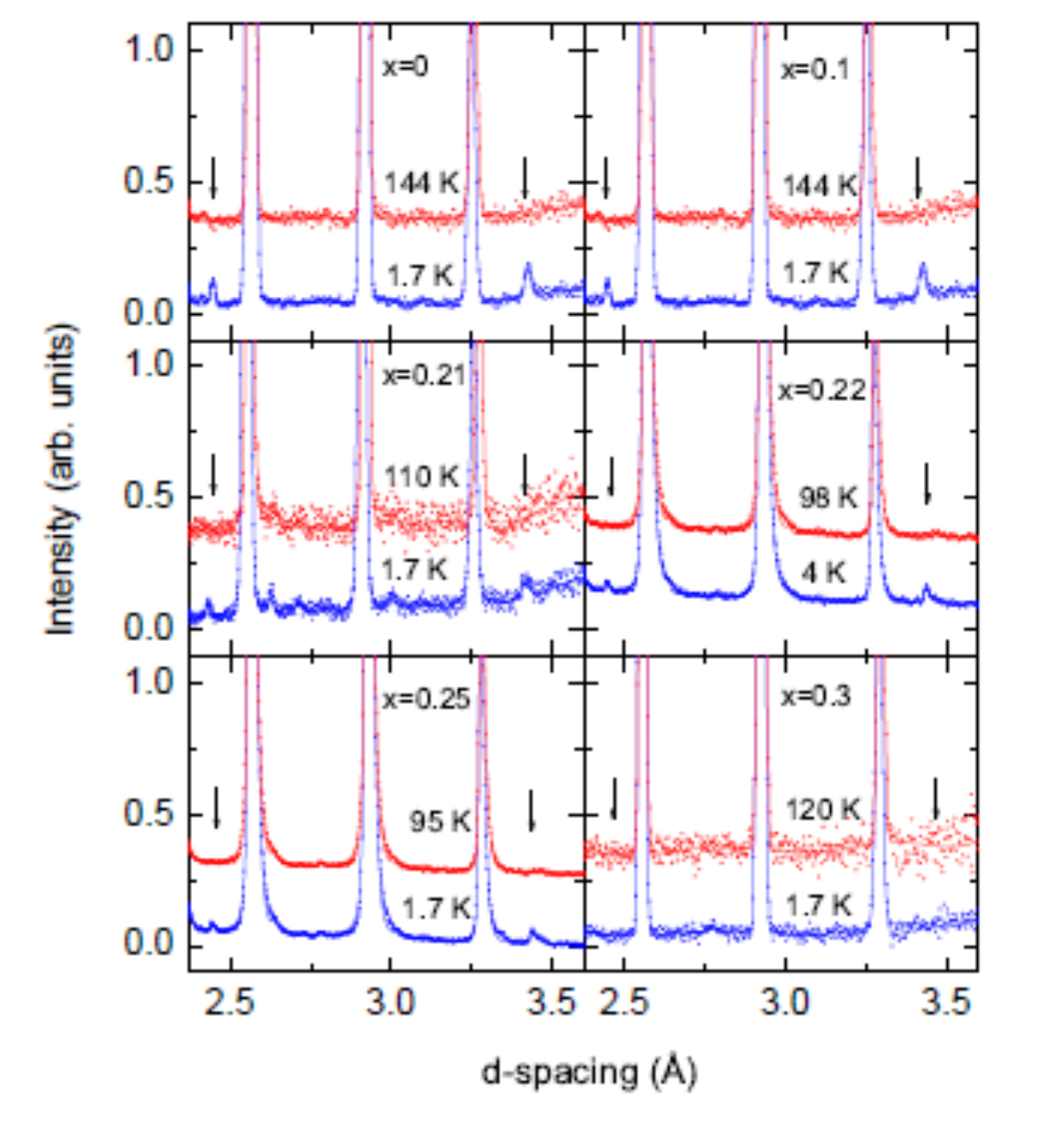}
\vspace{-0.15in}
\caption{(Color online) Neutron diffraction at 1.7\,K with the magnetic Bragg peaks at a $d$-spacing of 2.45\,\AA and 3.43\,\AA indicated by the arrows. They are absent above $T_N$, for $x=0$, 0.1, 0.21, 0.22 and 0.25. At $x=0.3$, no magnetic peaks are observed.
\label{Fig7} }
\vspace{-0.2in}
\end{figure}

As a further check on our data, the values of $T_N$ determined by neutron diffraction are in excellent agreement with those determined by peaks in the temperature derivative of the magnetization (Fig. 8 and Table 1). These magnetization peaks decrease in magnitude because of the progressive attenuation of the magnetic signal due to increasing K content until they are no longer detected for $x\geq0.21$.

A full analysis of the magnetic structure was performed using the allowed subgroup magnetic symmetries of $Fmmm$. All possible models were tested but only the magnetic space group $F_{c}mm^{\prime}m^{\prime}$ resulted in a proper fit to the data. Removal of the time reversal symmetry from two of the mirror planes resulted in an antiferromagnetic arrangement of the magnetic moments with a magnetic wave vector Q = (1,0,1); that is, the Fe magnetic moments are antiferromagnetically coupled in the $x$ and $z$ directions and ferromagnetically coupled along the $y$ axis. This model is consistent with similar results previously reported for the parent BaFe$_2$As$_2$ material [15,16,44]. 

\begin{figure}[!t]
\centering
\includegraphics[width=\columnwidth]{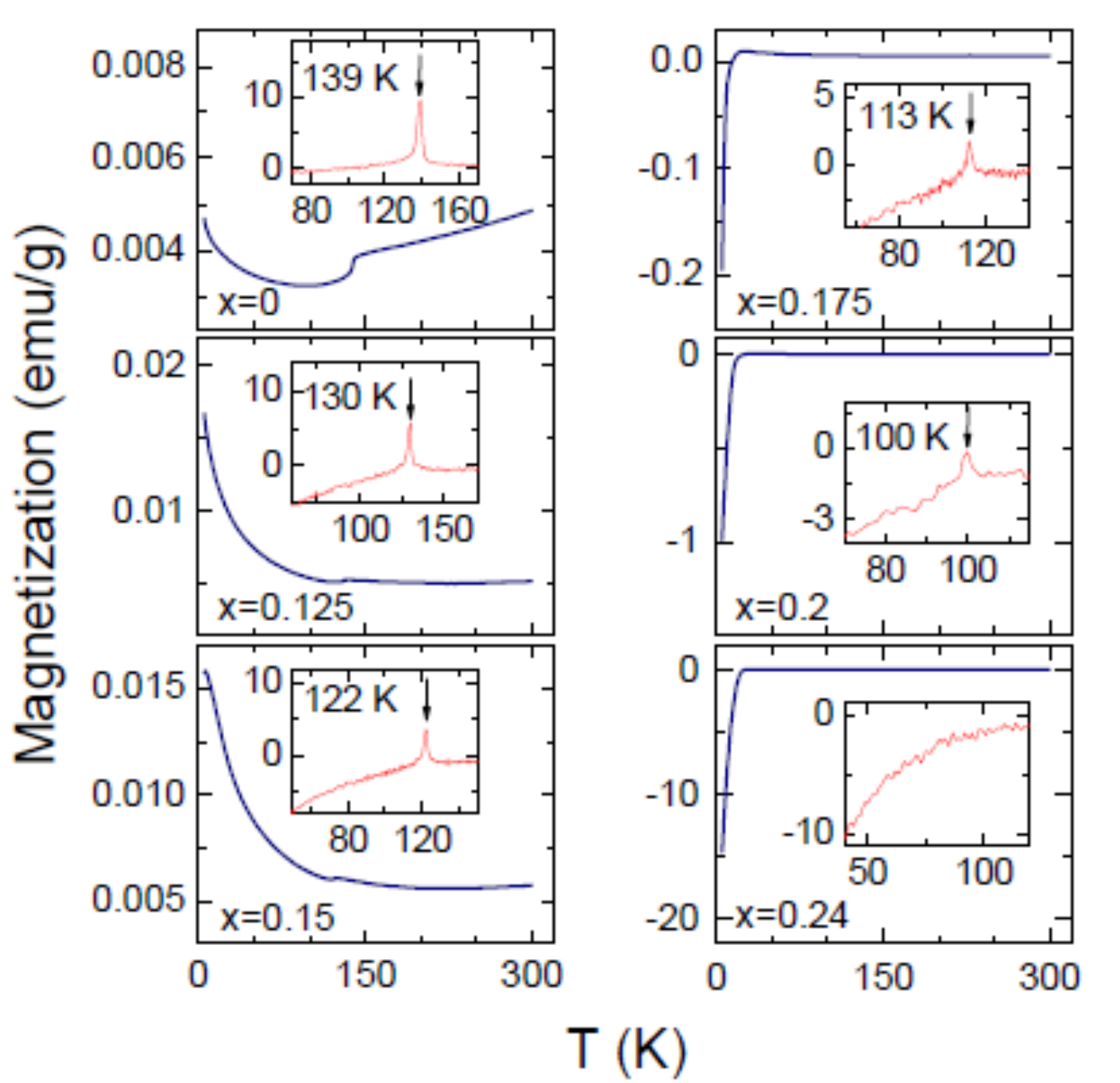}
\vspace{-0.15in}
\caption{(Color online) SQUID magnetization measurements for $x=0$, 0.125, 0.15, 0.175, 0.2 and 0.24 in 2\,kG applied magnetic field. Insets are the first derivatives of the magnetization curves, d$M$/d$T$ (10$^{-5}$), used to determine the N\'eel temperatures given by the arrows. For $x>0.2$, the magnetization anomaly at $T_N$ is too weak to be detected. 
\label{Fig8} }
\vspace{-0.2in}
\end{figure}

\begin{figure}[!b]
\centering
\includegraphics[width=\columnwidth]{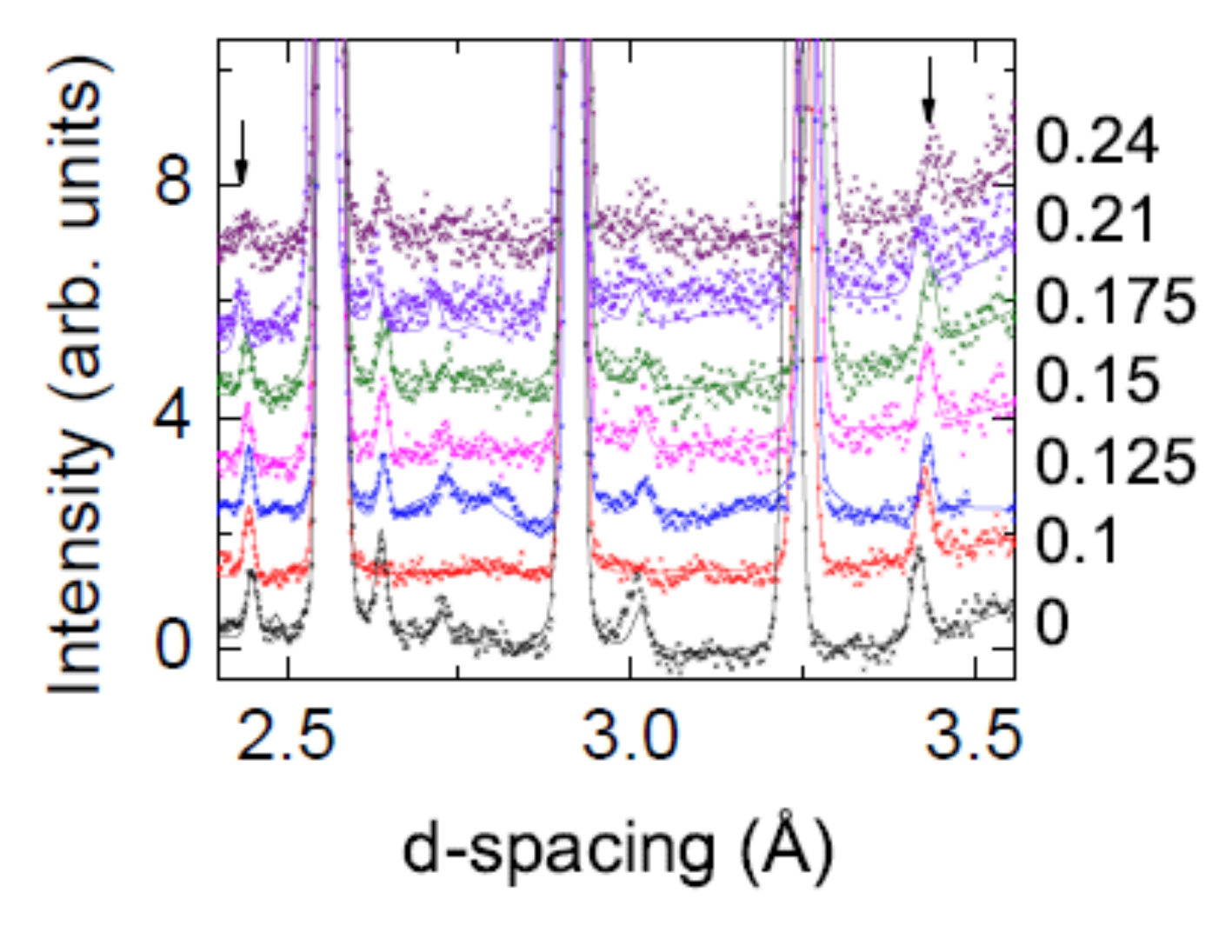}
\vspace{-0.15in}
\caption{(Color online) Dependence of the neutron diffraction intensity for Ba$_{1-x}$K$_x$Fe$_2$As$_2$ at 1.7\,K with $x$. The magnetic Bragg peaks are shown by the arrows. The solid lines represent the calculated intensity of the Rietveld refinement.
\label{Fig9} }
\vspace{-0.2in}
\end{figure}

Rietveld refinements of both the atomic and magnetic structures were performed simultaneously as a function of temperature and doping, allowing the magnetic moment to be defined in absolute units by normalization of the intensity of the magnetic Bragg peaks to the structural Bragg peaks. The neutron data displayed in Fig. 9 were collected at 1.7\,K and normalized to the sample mass and exposure time (measured in beam pulses). The figure qualitatively shows the intensities of the magnetic (121) and (103) Bragg reflections to remain roughly unchanged for the $x=0$ and 0.1 samples followed by a monotonic decrease upon increasing the K content until they nearly vanish at $x=0.24$. The refinements show that the magnetic moment drops from $\mu=0.75\mu_B$ for the parent BaFe$_2$As$_2$ material to $0.46\mu_B$ for $x\sim0.25$ (see Table 2). No magnetic peaks are observed beyond this value. 

\begin{figure}[!t]
\centering
\includegraphics[width=\columnwidth]{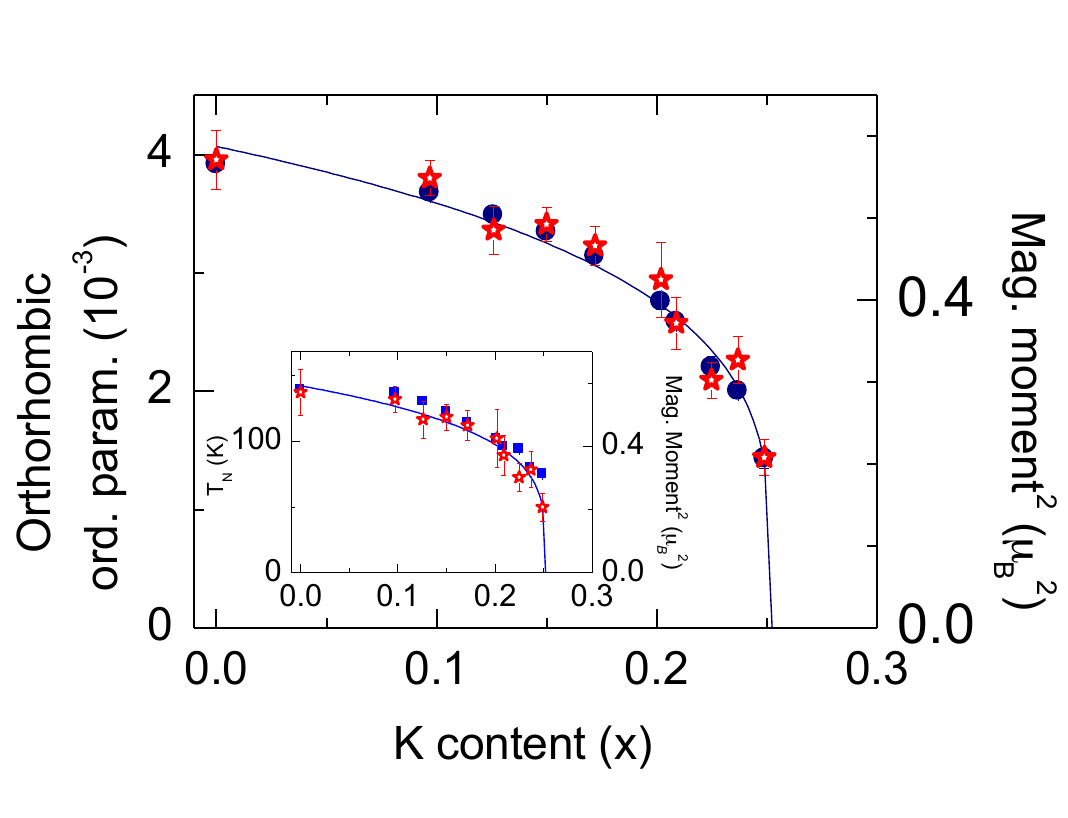}
\vspace{-0.15in}
\caption{(Color online) Dependence of the square of the magnetic moment, $\mu^2$ (red stars), and the orthorhombic order parameter, $\delta=(a-b)/(a+b)$ (blue circles), at 1.7\,K on the potassium concentration, $x$. The inset shows a comparison of $T_N$ and $\mu^2$ \textit{vs} $x$, showing that $T_N\propto \mu^2$. Solid lines in the main panel and the inset are a fit to $(1-x/x_c)^{2\beta}$ with $\beta=0.125$.
\label{Fig10} }
\vspace{-0.2in}
\end{figure}

The doping dependence of the two order parameters making up the AF/O phase is shown in Fig. 10. We compare $\delta$ and $\mu^2$ \textit{vs} $x$, showing that they are directly proportional over the entire range of AF/O order. We have not been able to measure any samples between  $0.25\leq x\leq0.28$, but a power law fit to $\delta$ and $\mu^2$ close to $x_c$, \textit{i.e.}, to $(1-x/x_c)^{2\beta}$, gives a critical concentration of 0.252 with an exponent of $\beta=0.125(1)$ for the structural and magnetic order parameters. The inset to Fig. 10 shows that $T_N$ is also proportional to $\mu^2$. In a mean-field model, $T_N$ scales as $J\mu^2$, where $J$ is the effective interionic exchange interaction, so this result would seem to indicate that $J$ is approximately independent of $x$ over this range.

\subsection{Superconductivity and Phase Coexistence}

\begin{figure}[!htb]
\centering
\includegraphics[width=0.6\columnwidth]{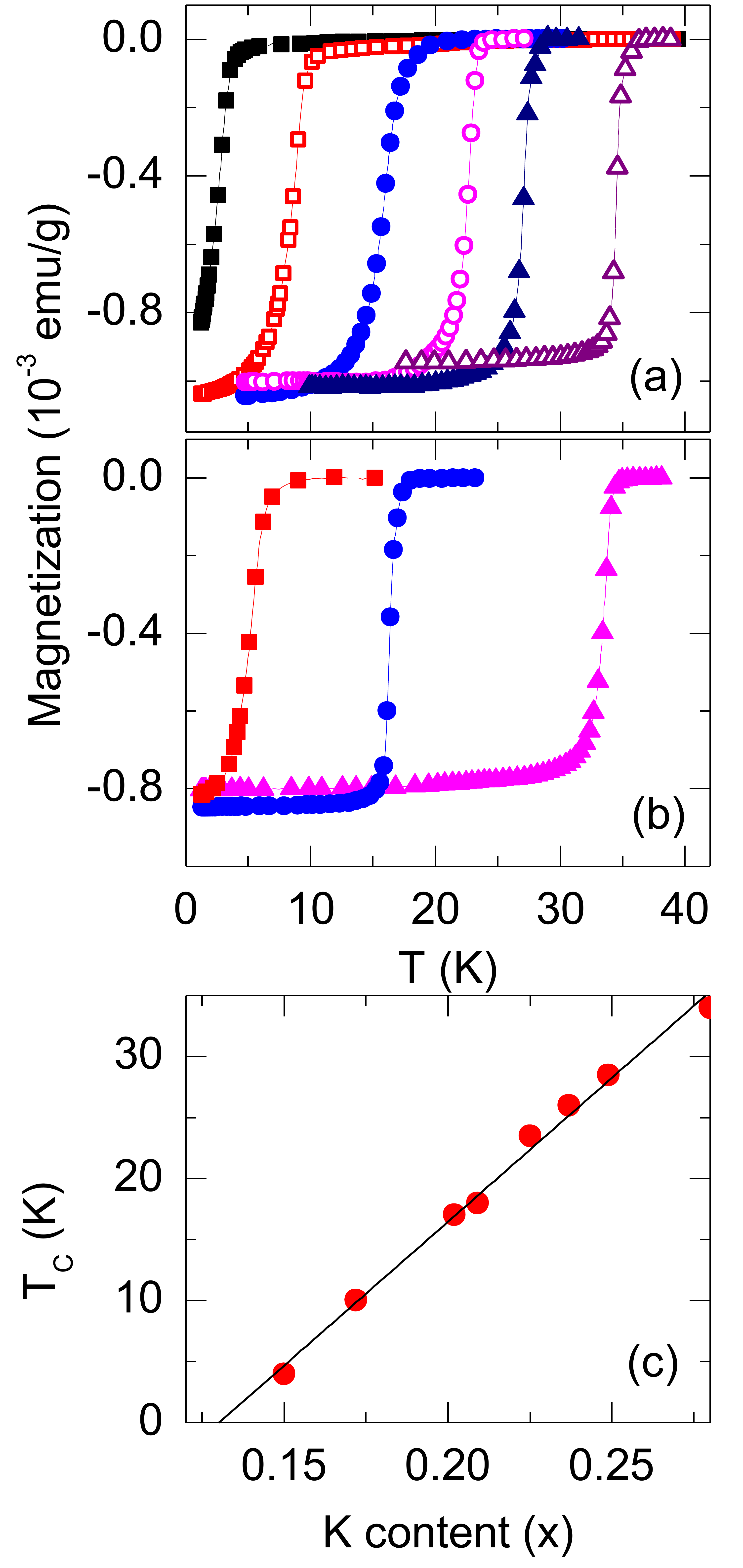}
\vspace{-0.15in}
\caption{(Color online) SQUID magnetization measurements (zero field-cooled) in 0.1\,G magnetic field for (a) $x=0.15$ (solid squares), 0.175 (open squares), 0.21 (solid circles), 0.22 (open circles), 0.25 (solid triangles), $x=0.3$ (open triangles), and (b) 0.5 (solid triangles), 0.7 (solid circles), 0.9 (solid squares) showing well-defined superconducting transitions. Magnetization values are normalized to mass of the samples. (c) Superconducting transition temperatures (onset $T_c$) of the underdoped compounds. Solid line represents the linear regression showing that the critical concentration for superconductivity is 0.130(3). 
\label{Fig11} }
\vspace{-0.2in}
\end{figure}

Bulk superconductivity is observed for all samples with $x\geq0.15$. Zero-field magnetization data shows that superconducting transition temperatures peaks at $\sim38$\,K for $x=0.4$ before it slowly decreases to 3\,K for the end member KFe$_2$As$_2$ (Fig. 11). By comparing the samples' magnetic moment with that of a Sn-powder sample of similar volume, we estimate that these samples are bulk superconductors with a volume fraction of at least 80\%. The uncertainty is due to variations in the demagnetization factor between samples. All the superconducting transitions are well-defined and sharp, even close to the critical concentration where $T_c$ is varying rapidly with $x$. Even low levels of compositional inhomogeneity associated with the uneven distribution of Ba/K ions would be revealed in magnetic susceptibility measurements by a broad or stepped-like transition from the normal state to the superconducting state, so this is further evidence of the sample quality. The increase in $T_c$ at low doping is approximately linear with dopant concentration up to $x\sim0.25$, so we have estimated the critical concentration for the onset of superconductivity using linear regression to be $x_c=0.130(3)$. 

In Ref.  [6], we discussed the behavior of the order parameters below $T_c$. We observed a small reduction in both the magnetic and structural order parameters of approximately 5\% at $x=0.21$ and 0.24, without seeing evidence of additional phases. In a scenario in which the sample divides into separate mesoscopic regions of AF/O phase and superconducting phase, this would imply that 95\% of the sample remains in the AF/O phase and only 5\% becomes superconducting. This is inconsistent with the magnetization measurements showing bulk superonductivity. While it is not possible to rule out the presence of other phases, the results indicate that there is microscopic phase coexistence of magnetism and superconductivity, with the reduction in the AF/O order parameters being due to a competition with the superconducting order parameter. This competition has been discussed extensively by Fernandes \textit{et al} [45], who show that there should be an additional phase boundary, with a positive slope \textit{vs} $x$, between the coexistence region and the region of purely superconducting phase. We have not yet identified any anomalies corresponding to the phase line below $T_c$, so we assume that it rises steeply with $x$.

\section{Discussion}
The overall phase diagram of Ba$_{1-x}$K$_x$Fe$_2$As$_2$ is shown in Fig. 12. We first discuss the nature of the spin-density wave order and orthorhombic order. Unlike the electron-doped compounds, where the two transitions split within increased doping, the two transitions are coincident and first-order in Ba$_{1-x}$K$_x$Fe$_2$As$_2$. We reported the first-order character of the transition by the observation of volume anomalies at $T_s$ [6]. Similar volume anomalies were also observed by Tegel \textit{et al} [46] in unsubstituted SrFe$_2$As$_2$ and EuFe$_2$As$_2$ but, because of the small magnitude of these anomalies, the authors suggested that the structural phase transition may be second-order. However, other authors reported first-order transitions in polycrystalline SrFe$_2$As$_2$ [20] and single crystals of CaFe$_2$As$_2$ [21] and BaFe$_2$As$_2$ [44]. In the latter reference, a first-order-like hysteresis was obtained for the intensities of the (101) Bragg peak when measured on cooling and warming. However, no such hysteresis was observed by Wilson \textit{et al} [27], when examining their BaFe$_2$As$_2$ single crystal. The systematic observation of volume anomalies across the phase diagram is unambiguous evidence that, at least in this system, all the transitions are first-order, although weakly first-order with extremely small hysteresis.

\begin{figure}[!t]
\centering
\includegraphics[width=\columnwidth]{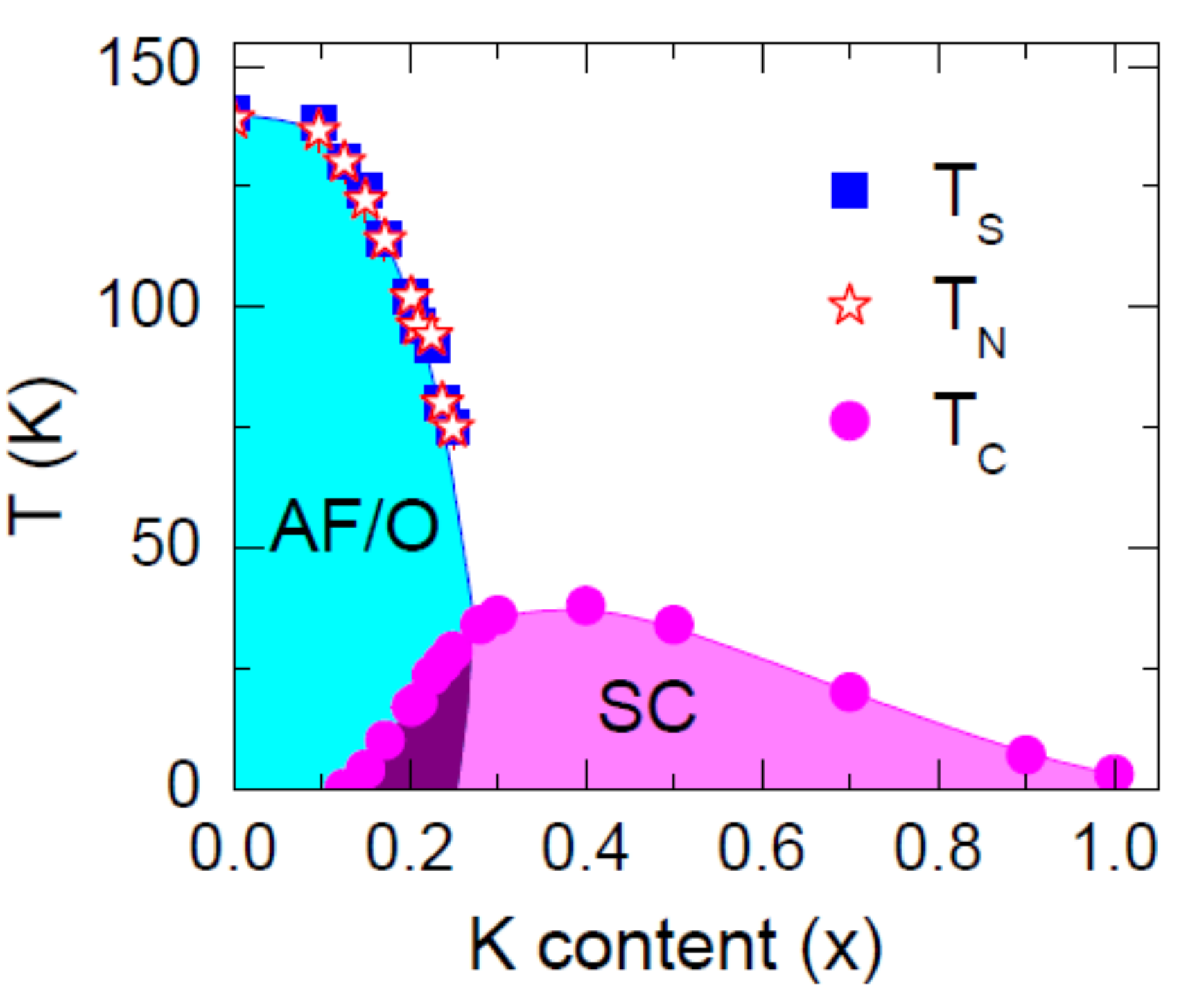}
\caption{(Color online) Phase diagram of Ba$_{1-x}$K$_x$Fe$_2$As$_2$ with the superconducting critical temperatures, $T_c$ (circles),  the N\'eel temperatures, $T_N$ (stars), and the structural transition temperatures, $T_s$ (squares).
\label{Fig12} }
\vspace{-0.2in}
\end{figure}

Phenomenological theory of magnetoelastic coupling predicts the possibility of simultaneous first-order transitions that are driven by a linear-quadratic term in the Ginzburg-Landau expansion [22,23] (see also Ref.  [47]). This is the lowest-order term allowed by symmetry. If the magnetic transition were to occur at higher temperature than the structural transition (in the absence of any competition), the magnetic order would drive the structural order in a simultaneous first-order transition. The converse would produce two split transitions as seen in most of the iron-based compounds [24]. There is a report of a split transition in Ba$_{1-x}$K$_x$Fe$_2$As$_2$ based on nuclear magnetic resonance results [25], but this is a local probe, which cannot necessarily identify compositional fluctuations. As we have already discussed, the neutron measurements, which represent true averages over the bulk, are quite unambiguous that the two transitions are simultaneous, although there could be some rounding of the transitions at higher doping from small compositional fluctuations. 

The two order parameters are directly proportional to each other as a function of temperature, which seems to indicate an unusual biquadratic coupling in the Ginzburg-Landau expansion, rather than a linear-quadratic coupling. This is usually only observed at a tetracritical point [26,27] where two phase boundaries intersect, whereas our observations extend over a range of compositions. There are a number of possible reasons for this. The most intriguing and exotic idea is that the AF/O order parameters are both secondary to another order parameter and directly driven by it. This would be the case in, for example, valley density wave theory, in which a mother density-wave drives both the magnetic and charge-density-wave orders [48]. A second explanation is provided by the recent theoretical work of A. Nevidomskyy [49], which uses a microscopic Kugel-Khomskii model to produce a biquadratic spin-orbital term in the free energy. A subtle, but ultimately more conventional explanation, is that the coupling is linear-quadratic after all, but the proximity to a first-order transition produces a temperature dependence that is approximately equivalent to biquadratic coupling to first order [29]. This is in the context of a theory in which Ising-nematic order, produced by an itinerant model of Fermi surface nesting, drives the structural transition. Support for this explanation is provided by Fig. 10, where the doping dependence of the magnetic and structural order parameters at low temperature indicates a linear-quadratic coupling. Whatever the eventual explanation, it is clear that this result is key to understanding the nature of the normal state and the role of nematic order in the eventual superconductivity.

The strong coupling between AF/O order parameters persists into the regime of phase coexistence with superconductivity. We have already argued that Ba$_{1-x}$K$_x$Fe$_2$As$_2$ is characterized by microscopic phase coexistence because mesoscopic phase separation would result in a significant decrease in the volume fraction of the AF/O phase below $T_c$. The consensus in favor of microscopic phase coexistence has existed for some time in the electron-doped superconductors, such as BaFe$_{2-x}$Co$_x$As$_2$ [24], where the phase boundary below $T_c$ to a non-magnetic, purely superconducting region has also been identified. Theoretically, this behavior is consistent with unconventional $s_{\pm}$ pairing of the Cooper pairs suggesting that itinerant long range magnetism and superconductivity may coexist and compete for the same electrons [45]. However, the idea of microscopic phase coexistence was more controversial in Ba$_{1-x}$K$_x$Fe$_2$As$_2$  because of local probe measurements that seemed to indicate a phase separation into mesoscopic regions of magnetism and superconductivity [30,31]. Since the most recent $\mu$SR data are also consistent with microscopic phase coexistence [32], it appears that the earlier reports may have been due to compositional fluctuations close to the phase boundaries and that microscopic phase coexistence has now been confirmed.

\begin{figure}[!t]
\centering
\includegraphics[width=\columnwidth]{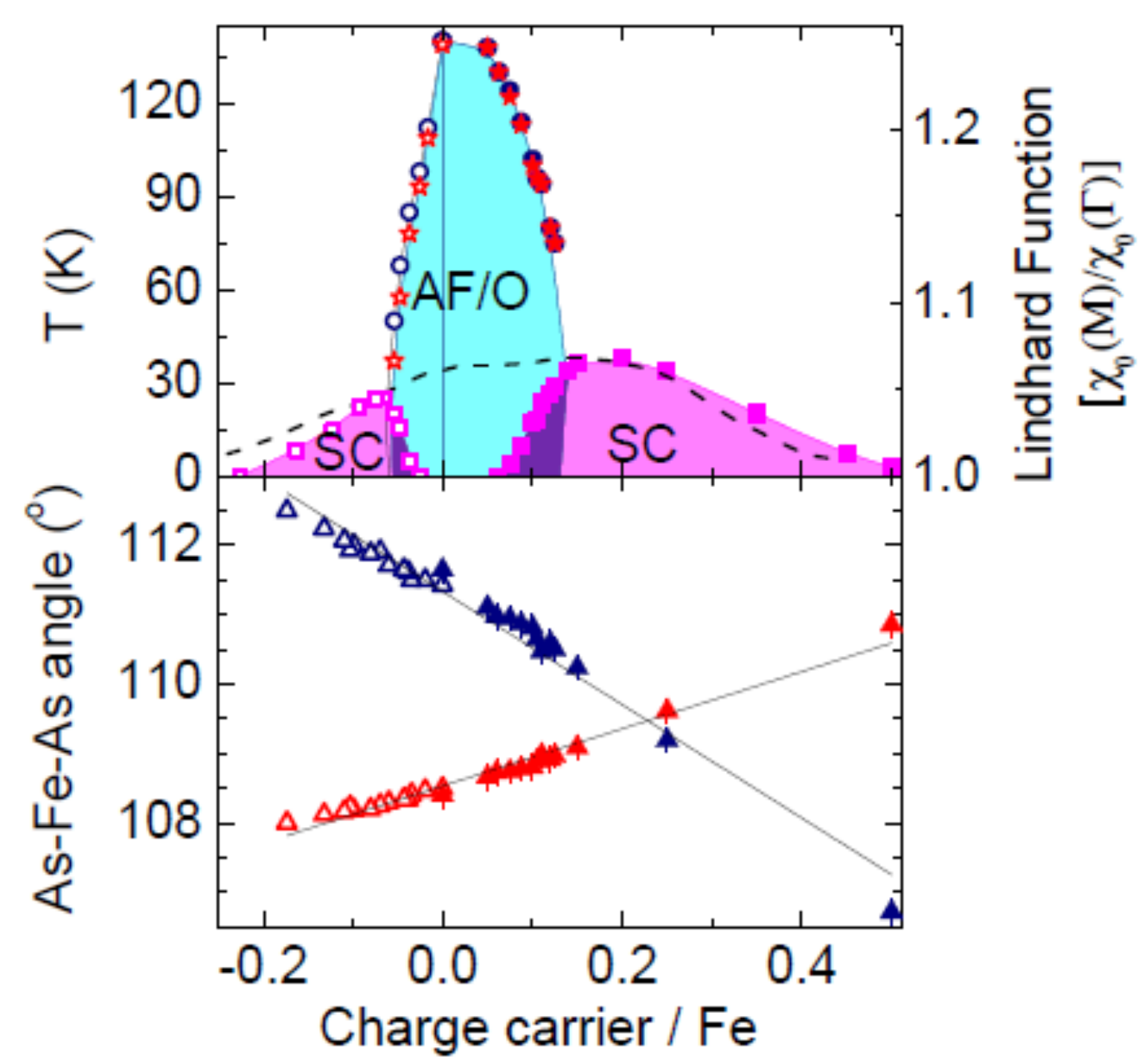}
\vspace{-0.15in}
\caption{(Color online) (Top panel) Magnetic and structural phase diagram of electron-doped Ba(Fe$_{1-x}$Co$_x$)$_2$As$_2$ and hole-doped Ba$_{1-x}$K$_x$Fe$_2$As$_2$ with the superconducting critical temperatures, $T_c$ (squares), N\'eel temperatures, $T_N$ (stars) and structural transition temperatures, $T_s$ (circles). The $x$-axis is normalized to the charge carrier per iron atom. Data for the electron-doped side where the transition temperatures are represented with open symbols are taken from Ref  [50]. The error bars for $T_N$ and $T_s$ values in the hole-doped side are within the symbols. The dashed line enveloping the superconducting dome represents the Lindhard function taken from Ref  [33]. (Bottom panel) Charge carrier dependence of the As-Fe-As bond angles for both electron- and hole-doping. Solid triangles represent the results of our neutron diffraction study at 1.7\,K for the hole-doped Ba$_{1-x}$K$_x$Fe$_2$As$_2$. At this temperature one of the As-Fe-As angles splits due to orthorhombic distortion below $x=0.3$. Therefore, we took the average of these two splitting angles. The As-Fe-As bond angle data for the electron doped side is taken from Ref  [51]. Solid lines are guide to the eye.
\label{Fig13} }
\vspace{-0.2in}
\end{figure}

Finally, we discuss the electron-hole asymmetry in the phase diagram, shown in Fig. 13, where we have added data from the literature [50,51] to allow a comparison with the more commonly studied electron-doped superconductors. In this phase diagram, the $x$-axis is normalized to the number of charge carriers per Fe atom. Neupane \textit{et al} have recently suggested that this asymmetry is due to differences in the effective masses of the hole and electron pockets [33]. This is justified by ARPES data that show that hole doping can be well described within a rigid band approximation [52]. An \textit{ab initio} calculation of the Lindhard function of the non-interacting susceptibility at the Fermi surface nesting wavevector shows exactly this asymmetry, with a peak at $x\sim0.4$ where the maximum $T_c$ occurs. Our recent inelastic neutron scattering measurements of the resonant spin excitations that are also sensitive to Fermi surface nesting have shown a similar correlation between the strength of superconductivity and the mismatch in the hole and electron Fermi surface volumes [34], that is responsible for the fall of the Lindhard function at high $x$.  An overall envelope may be drawn (dashed line in Fig. 13) to encompass both the hole and electron superconducting domes of the phase diagram. If anything, the Lindhard function underestimates the asymmetry, predicting a larger superconducting dome on the electron-doped side. We attribute this behavior to the fact that the iron arsenide layers remain intact in the potassium substituted series, whereas Co substitution for Fe disturbs the contiguity of the FeAs$_4$ tetrahedra and interferes with superconductivity in these layers.

Interestingly, the maximum overall $T_c$ also correlates with the perfect tetrahedral angle of $\sim109.5^\circ$ as demonstrated in the bottom panel of Fig. 13. In the plot, average $<$As-Fe-As$>$ bond angles for our K-substituted series have been extracted from the Rietveld refinements. The As-Fe-As bond angles for BaFe$_{2-x}$Co$_x$As$_2$  are extracted from the literature [51]. The continuity of the bond angles across the electron-doped and hole-doped sides of the phase diagram is remarkable and the crossing of the two independent angles at $x\sim0.4$ to yield a perfect tetrahedron and maximum $T_c$ is clear. This has been remarked before in other systems [35,53]. It is possible that these two apparently distinct explanations for the maximum $T_c$ are two sides of the same coin. In a theoretical analysis of the 1111 compounds [38], it has been suggested that the pnictogen height is important in controlling the energies of different orbital contributions to the $d$-bands and so affect the strength of the interband scattering that produces superconductivity.

We now turn our attention to the SDW region of the phase diagram. While it is clear that spin-density-wave order has to be suppressed in order to allow superconductivity to develop, it is not immediately clear what is responsible for the suppression. Both the strength of magnetic interactions and superconductivity, at least in an itinerant model, depend on the same Lindhard function [54], the former on the peak in the susceptibility at the magnetic wavevector, and the latter on an integral over the Fermi surfaces. It would seem therefore that the magnetic transition temperature should also peak at $x\sim0.4$. One intriguing reason why it would peak at $x=0$ is because magnetic order is more sensitive to disorder-induced suppression of the peak susceptibility whereas superconductivity is more robust. There is some support for this idea from the observation that isoelectronic doping produces a similar suppression of magnetic order than seen with hole-doping [51]. On the other hand, Kimber \textit{et al} succeeded in rendering the parent BaFe$_2$As$_2$ material to exhibit zero resistance at 30.5\,K by the application of significant external pressures up to 5.5\,GPa [55], \textit{i.e.}, without introducing disorder, but they remarked that superconductivity needs to be confirmed by other bulk measurement techniques. Interestingly, the authors also correlate the induced $T_c$ with approaching a perfect tetrahedron angle of 109.5o similar to our observations for $x\sim0.4$.

In summary, we have synthesized high quality samples covering the full phase diagram of Ba$_{1-x}$K$_x$Fe$_2$As$_2$. Using high resolution neutron powder diffraction and SQUID magnetization measurements, we have investigated the effects of potassium substitution on superconductivity, structural transformation and magnetic ordering. Our measurements allowed the construction of a detailed magnetic and structural phase diagram, which displays a narrower phase coexistence region than the previous reports. Moreover, neutron diffraction and the SQUID magnetization data confirmed that magnetic and structural transitions are coincident with first order transitions. Additionally, we determined the effects of temperature and substitution on the various internal atomic and structural parameters. Our results confirm the importance of obtaining precise structural parameters across the whole phase diagram as a way of providing insight into the nature of the phase competition that underlies iron-based superconductivity. 

This work was supported by the Materials Sciences and Engineering Division of the Office of Basic Energy Sciences, Office of Science, U.S. Department of Energy, under contract No. DE-AC02-06CH11357.


\end{document}